\documentclass[aps,prb,superscriptaddress,twocolumn,floatfix]{revtex4-1}

\usepackage{graphicx,color}
\usepackage{dcolumn}
\usepackage{bm}
\usepackage{stmaryrd}
\usepackage{latexsym}
\usepackage{amssymb}
\usepackage{amsfonts}
\usepackage{amsmath}
\usepackage{subfigure}
\usepackage{verbatim}
\usepackage{multirow}

\begin{document}

\preprint{APS/123-QED}

\title{Electron-Nuclear Hyperfine Coupling in Quantum Kagome Antiferromagnets from First-Principles Calculation and a Reflection of the Defect Effect}
\author{Shunhong Zhang}
\affiliation{Institute for Advanced Study, Tsinghua University, Beijing 100084, China}
\author{Yi Zhou}
\affiliation{Beijing National Laboratory for Condensed Matter Physics, and Institute of Physics, Chinese Academy of Sciences, Beijing 100190, China}
\affiliation{CAS Center for Excellence in Topological Quantum Computation, University of Chinese Academy of Sciences, Beijing 100190, China}
\affiliation{Collaborative Innovation Center of Advanced Microstructures, Nanjing 210093, China}
\author{Feng Liu}
\affiliation{Department of Materials Science and Engineering, University of Utah, Salt Lake City, Utah 84112, USA}
\affiliation{Collaborative Innovation Center of Quantum Matter, Beijing 100084, China}
\author{Zheng Liu}
\email{zheng-liu@tsinghua.edu.cn}
\affiliation{Institute for Advanced Study, Tsinghua University, Beijing 100084, China}
\affiliation{Collaborative Innovation Center of Quantum Matter, Beijing 100084, China}
\date{\today}

\begin{abstract}
The discovery of ideal spin-1/2 kagome antiferromagnets Herbertsmithite and Zn-doped Barlowite represents a breakthrough in the quest for quantum spin liquids (QSLs), and  nuclear magnetic resonance (NMR) spectroscopy plays a prominent role in revealing the quantum paramagnetism in these compounds. However, interpretation of NMR data that is often masked by defects can be controversial. Here, we show that the most significant interaction strength for NMR, i.e. the hyperfine coupling (HFC) strength, can be reasonably reproduced by first-principles calculations for these proposed QSLs . Applying this method to a supercell containing Cu-Zn defects enables us to map out the variation and distribution of HFC at different nuclear sites. This predictive power is expected to bridge the missing link in the analysis of the low-temperature NMR data.
\end{abstract}
\pacs{Valid PACS appear here}

\maketitle

\section{Introduction} Quantum spin liquid (QSL)~\cite{Nature10QSL_review,RPP17QSL_review,RMP17QSL} is an emergent quantum phase in solid states that activates several fields of frontier physics, such as quantum magnetism, topological order~\cite{PRB1991Wen_MFT_spin_gap}, and high-temperature superconductivity~\cite{Science1987Anderson,RMP06dop_Mott}. For decades, kagome antiferromagnets have been intensively searched and studied as candidates to realize the QSL state~\cite{RMP16QSL,Science08Lee}. Significant advances have been made, with synthetic Herbertsmithite [Cu$_3$Zn(OH)$_6$Cl$_2$] as a prototypical example~\cite{JACS05synthesis_HS}. Recently, first-principles calculations suggested Zn-doped Barlowite [Cu$_3$Zn(OH)$_6$FBr] as a sibling QSL candidate~\cite{PRB15DFT_barlowite} and subsequent experiments observed promising signals~\cite{CPL17NMR_barlowite,PRM18Cu4_single_crystal}. The two compounds share a similar layered kagome spin lattice formed by S=1/2 Cu$^{2+}$ ions. The absence of long-range magnetic order in these kagome magnets, a feature of QSL, preserves down to several tens of mK, despite the fact that the primary nearest neighbor (NN) antiferromagnetic (AFM) interaction is of the order of 10$^2$ K~\cite{CPL17NMR_barlowite,JACS05synthesis_HS,PRM18Cu4_single_crystal}.

\begin{figure*}
\centering
\includegraphics[width=15cm]{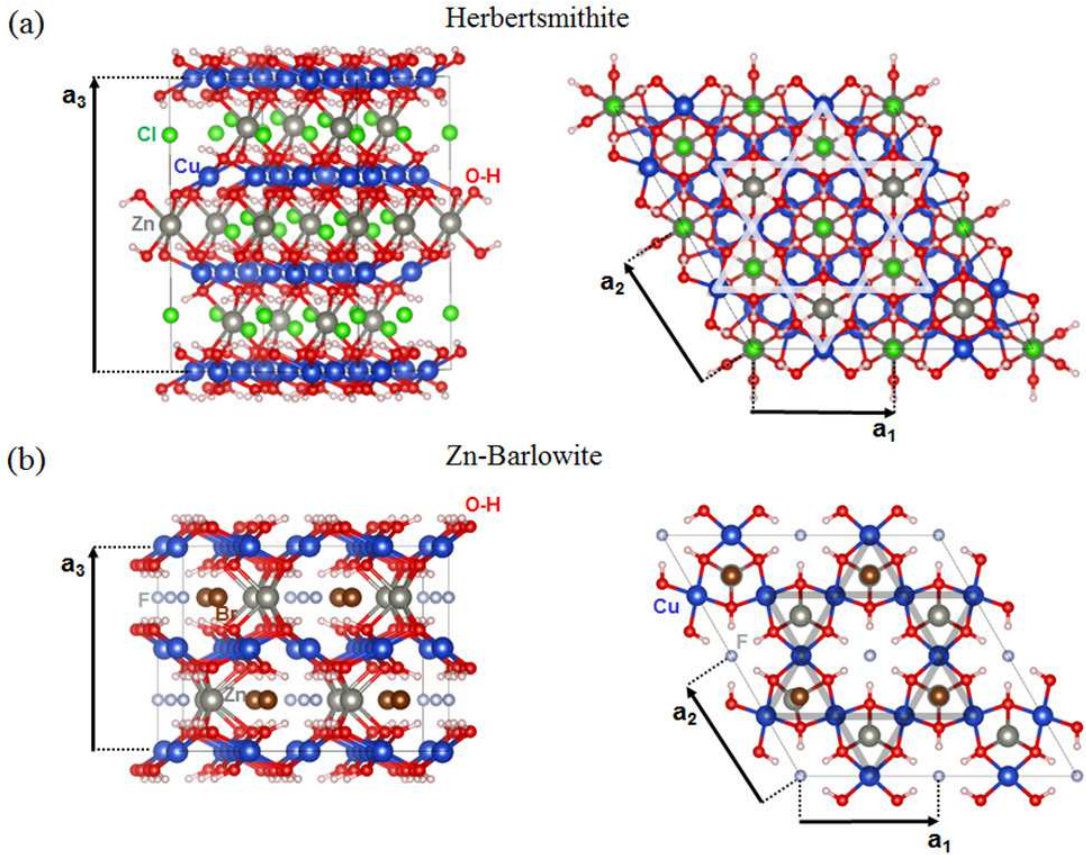}
\caption{Crystal structures of (a) Herbertsmithite and (b) Zn-Barlowite.  The kagome geometry of the Cu$^{2+}$ ions is indicated by shaded Stars of David in the top view. The three lattice vectors are marked by $\textbf{a}_{i=1,2,3}$.}
\label{struct}
\end{figure*}

Experimentally, one severe obstacle to clarifying the nature of the QSL ground state is the defect spin dynamics~\cite{CRP16Mendel}. For example, macroscopic susceptibility measurements on both Herbertsmithite~\cite{JACS05synthesis_HS} and Zn-Barlowite~\cite{CPL17NMR_barlowite} revealed Curie-like divergence at the low temperature overwhelming the intrinsic kagome AFM susceptibility, which is perceived as a consequence of isolated spins from extra Cu$^{2+}$ ions between the kagome layers.

Nuclear magnetic resonance (NMR) spectroscopy has played a prominent role in detecting the magnetic order in solids. By detecting the nuclear resonance peaks in presence of an external magnetic field, NMR probes the spin susceptibility via the hyperfine coupling (HFC) between the nuclear spin and the neighboring electronic spins. Since HFC is of short-range, it renders a nuclear site resolution, making it possible to differentiate the resonance of  a nucleus in a local environment free of defect from the one situated in a defect-contaminated local environment. One common feature revealed by different NMR measurements is that the intrinsic spin susceptibility in these kagome compounds decreases rapidly at low temperatures, in striking contrast to the macroscopic susceptibility. With the persistent improvement of sample qualities, NMR is now possible to address the fundamental theoretical questions  - ``Is the ground state gapped or gapless?'' and ``What is the quantum number of the excitations?''

Active investigations are underway to reach a consensus of opinion. One key issue is how to subtract the  background defect signal accurately at low temperatures. Typically, the NMR peaks are convoluted by defect signals when the temperature goes down, making it tricky to naively identify the peak positions. It is worth noting that the smeared peaks in NMR spectra may come from random spin configurations, such as spin glass or valence bond glass, as well \cite{RMP17QSL}. 

The motivation of this Article is to extend the application of the HFC calculation method as established for conventional magnetic systems~\cite{1968PRLBe_HFC_Jena,PRB2006HFC_Gaussian_APW,PRB06HFC_Kawazoe,MRC2010Review_NMR} to QSL candidates. We note that since (spin) density functional theory (DFT) is practically a single-electron mean-field approach, it is traditionally perceived that hardly can any useful QSL information be inferred from this method. Interestingly, we realize that HFC represents an ideal property for DFT analysis without explicitly constructing the QSL wavefunction. After a brief review of the basic aspects of NMR and HFC in Sec. II, we will discuss in details the physical considerations and propose the practical recipe in Sec. III. We apply this recipe to Herbertsmithite and Zn-Barlowite with the computation details described in Sec. IV, and in Sec. V we demonstrate that such calculated HFC constants indeed achieve good consistency with experimental results. In Sec. VI, by purposely introducing different types of defects, we can further map out the variation of HFC at different nuclear sites [Fig.~\ref{fig-1}(b)], which bridges the missing link in the analysis of  the low-temperature NMR data.  Sec. VII concludes this Article.

\section{Basic aspects of NMR and HFC}
A free nuclear spin in an external magnetic field ($B_{ext}$) exhibits a resonance frequency ($f_0$) 
due to the Zeeman effect. Embedded in a material,  the nuclear resonance frequency will shift to $f$ due to the presence of internal magnetic field ($B_{int}$) induced by the electron spins. For example, Cl  in Herbertsmithite, F in Zn-Barlowite and O in both compounds possess stable isotopes $^{35}$Cl, $^{19}$F and $^{17}$O with nuclear spin I = 3/2, 1/2, and 5/2,  respectively, which have certain resonance frequencies. On the other hand,  the electron spins in both compounds localize on the Cu$^{2+}$ ions, in the ideal case lying in the kagome planes (See Fig.~\ref{struct}), which generates hyperfine field.  We note that since QSL candidates are typically Mott (or charge-transfer) insulators, the chemical shift induced by orbital moment does not play an important role. In experiment, the chemical shift is commonly subtracted from the data as a temperature-independent constant.

The ratio between $B_{int}$ and the expectation value of the electron spin $\langle S \rangle$ is defined as the HFC constant, i.e.
\begin{eqnarray} \label{Bint}
B_{int}^I=A_{hf}^I\langle S\rangle,
\end{eqnarray}
in which we add a label $I$ to denote different nuclear probes. We will focus on powder sample measurements, so the scalars in Eq.~(\ref{Bint}) are considered to be implicitly orientation averaged. 

From the NMR shift, one can extract the information on the static electron spin susceptibility ($\chi_S$):
\begin{eqnarray}\label{eq:K}
K^I\equiv\frac{f}{f_0}-1=\frac{B_{int}^I}{B_{ext}}=A_{hf}^I\chi_S(T)
\end{eqnarray}
Meanwhile, the static spin susceptibility can also be measured by macroscopic magnetometric methods, which we denote as $\chi_S^{macro}$. Whenever $K$ and $\chi_S^{macro}$ display a good linear correlation, a common practice to determine $A_{hf}$ experimentally is via the $K$-$\chi_S^{macro}$ slope. 

Microscopically, $B_{int}$ has two origins~\cite{NMR_Principles,PRB1987Nickel_HF}. A dipole field that captures the long-range anisotropic interaction:
\begin{equation}
\label{dipolar}
(\textbf{B}_\textrm{int}^{dipole})_{ij}=\frac{ \mu_0 g_e \mu_B }{4\pi}\int{\frac{\rho_s({\bf{r}}+{\bf{R}}_I)}{r^3}\frac{3r_ir_j-\delta_{ij}r^2}{r^2}d\bf{r}},
\end{equation}
and the Fermi contact term that takes care of the non-vanishing unpaired electron wavefunction inside the nucleus:
\begin{equation}\label{fermi}
B_\textrm{int}^{Fermi}=\frac{2\mu_0 g_e \mu_B }{3}\rho_s({\bf{R}}_I),
\end{equation}
in which $\mu_0$ is the vacuum  permeability, $g_e$ is the electron g-factor, $\mu_B$ is the Bohr magnetic moment, $\rho_s$ is the electron spin density, and $\textbf{r}$ and $\textbf{R}_I$ denote the coordination of electrons and nuclei, respectively.  

For a powder sample, the crystal orientation is random and the external magnetic field is expected to have an equal probability to align with the three principle axes of  $(\textbf{B}_\textrm{int}^{dipole})_{ij}$. We thus evaluate $B_{hf}$ in Eq. (\ref{Bint}) as the arithmetic average along the three principle directions:
\begin{eqnarray}\label{averageB}
B_{int}^I=B_\textrm{int}^{Fermi}+Tr[(\textbf{B}_\textrm{int}^{dipole})_{ij}]/3=B_\textrm{int}^{Fermi}
\end{eqnarray}
Note that the matrix $(\textbf{B}_\textrm{int}^{dipole})_{ij}$ is traceless. Under this approximation, the NMR peak frequency is dictated by the Fermi contact term, while the dipolar term contributes to peak broadening.

The key information to calculate $A_{hf}$ is nothing but $\rho_s(\textbf{r})$. Note that $\langle S\rangle=\frac{1}{\textit{N}}\int \rho_s(\textbf{r})d\textbf{r}$, where \textit{N} is a normalizing factor, depending on whether $\langle S\rangle$ is averaged to one unit cell or one magnetic atom etc. Therefore, two of the three quantities in Eq.~(\ref{Bint}) are directly calculable given $\rho_s(\textbf{r})$, and $A_{hf}$ can be derived by taking the ratio.

\section{DFT-based HFC calculation}

\subsection{Established methodology for conventional magnetic systems}

Ground-state spin density is a calculable quantity exactly fit in DFT. Extensive efforts have been made to reliably reproduce $\rho_s$ in a variety of conventional magnetic systems, ranging from single atoms, clusters, metals and insulators \cite{PRB2006HFC_Gaussian_APW, PRB06HFC_Kawazoe, MRC2010Review_NMR, PRB03Ceder_NMR_TMO, PRL2011_YMnO_HFC}. An accurate representation of the core wavefunctions is the primary challenge. With the development of  augmented basis functions, projector-augmented-waves pseudopotentials, as well as exchange-correlation functionals, the most widely used DFT codes can already achieve essential reproducibility ~\cite{PRB03wien_HFC,PRB10wien_HFC}, which presents a solid basis for our generalization to QSLs.

To benchmark with experiment, the previous calculations typically start with the experimental spin structures, and $\rho_s$ is determined iteratively by minimizing the DFT energy functional without including an external magnetic field.  We should keep in mind that rigorously $\langle S \rangle$ in Eq.~(\ref{Bint}) is the expectation value of the electron spin under $B_{ext}$, instead of the ground-state expectation value. Therefore, such calculations only give the intrinsic hyperfine field, but ignore any $B_{ext}$-induced spin density change, e.g. polarization and canting. This issue is insignificant for a ferromagnet, in which $B_{ext}$ merely aligns the spins to a certain direction.  For an antiferromagnet, however, one should only apply the calculation results to nuclear probes that predominantly experience the intrinsic hyperfine field, e.g. the probe is the magnetic ion itself. In contrast, when the nuclear probe resides in the middle of two antiferromagentically coupled magnetic ions, the intrinsic hyperfine field largely cancels, whereas the HFC observed in experiment mainly arises from the ignored part. 

\subsection{Special considerations for QSL candidates}
The aforementioned problem becomes even more severe for a QSL that by definition consists of fluctuating spins down to 0K, i.e. without $B_{ext}$, $\langle S \rangle = 0$. It can be immediately seen that the intrinsic hyperfine field is exactly zero, and any measurable HFC arises from  $B_{ext}$-induced polarization. A natural attempt is to first calculate the response of the QSL ground state to  $B_{ext}$, and then do the HFC calculation based on that. However, this route is intimidating: not to mention how the QSL ground state can be described within DFT, it already costs tremendous efforts to calculate the susceptibility of QSL from simple effective models.

One interesting observation, as widely accepted in experiment, is that HFC in many cases is independent of temperature and magnetic field. If this perception is valid,  it is possible to construct a practical calculation recipe without dealing with the exotic QSL wavefunction. Intuitively, when the temperature is comparable to the primary spin exchange, the system is actually far from the quantum regime, exhibiting the classical Curie-Weiss behavior. So imagine that at a moderate temperature, we hypothetically turn up $B_{ext}$. Then $\langle S \rangle$ will gradually increase (no matter how), and ultimately saturate. The real-space spin density of this fully-polarized state can be calculated directly using the established DFT method.

In short, \textit{we propose performing DFT-based HFC calculation on QSL candidates by choosing a fully-polarized reference state, and the derived HFC constant is expected to be directly comparable to experiment, so long as this constant is independent of T and $B_{ext}$.}

We note that there are known examples, in which the HFC constant varies with T~\cite{JPSJ2008BaFeAs-NMR,Wu2019FeSe_NMR}. Therefore, in general one should examine first whether the application condition can be met. In the next subsection, we demonstrate that for Herbertsmithite and Zn-Barlowite the proposed calculation  can be put on a firmer footing. 

\subsection{Formal justification for Herbertsmithite and Zn-Barlowite}

In these two compounds, the electron spins arise from the Cu$^{2+}$ ions with a 3d$^9$ valence configuration. Under the local Cu-O square crystal field, four of the five 3d orbitals are fully occupied, leaving a single electron at the $d_{x^2-y^2}$ orbital on top.  We define the creation operator of this single electron as:
\begin{eqnarray}\label{cdag}
c_{i\sigma}^\dag=\int d\textbf{r}w(\textbf{r}-\textbf{R}_i)c_\sigma^\dag(\textbf{r}),
\end{eqnarray}
in which $i$ labels the Cu site and $\sigma=\uparrow,\downarrow$ labels the spin. Without loss of generality, the spin quantization direction is chosen to align with $B_{ext}$. The function $w(\textbf{r}-\textbf{R}_i)$ is the spatial wavefunction of this single electron, which predominantly consists of the $3d_{x^2-y^2}$ orbital, and should have certain hybridization with O and other more distant atoms. This hybridization results in nonvanishing $\rho_s$ around the nuclear region of these atoms, i.e. HFC. It is important to note that the function form of  $w$ does not depend on $i$ and $\sigma$.

We can formally write the many-body eigenstates of this system as:
\begin{eqnarray}
|\Phi^B_\alpha\rangle=f_\alpha^B(\{c_{i\sigma}^\dag\})|0\rangle,
\end{eqnarray}
in which $\alpha$ labels the eigenstates with the corresponding eigenenergy $E^B_\alpha$. The label $B$ reminds that an external magnetic field is present that affects both the eigenstates and the eigenenergy. All the complexities of QSL are wrapped into the abstract function  $f_\alpha^B(\{c_{i\sigma}^\dag\})$, which in general takes the form of products and superposition of $\{c_{i\sigma}^\dag\}$.

The expectation value of the electron spin $\langle S \rangle$ that appears in Eq.~(\ref{Bint}) can thus be calculated as:
\begin{eqnarray}\label{S}
\langle S \rangle = \frac{1}{Z}\sum_\alpha e^{-\beta E_{\alpha}}\langle \Phi_\alpha^B|c_{i\uparrow}^\dag c_{i\uparrow}-c_{i\downarrow}^\dag c_{i\downarrow} |\Phi_\alpha^B\rangle,
\end{eqnarray}
in which $Z=\sum_\alpha e^{-\beta E_{\alpha}}$ is the partition function. It is understood that in the absence $B_{ext}$,  the correct QSL eigenstates $|\Phi_\alpha^{B=0}\rangle$  should give $\langle S \rangle$=0. The equation demands that $|\Phi_\alpha^{B}\rangle$ maintains all the lattice symmetry. Therefore, all the Cu$^{2+}$ ions are equivalent, and $\langle S \rangle$ is independent of $i$.

The spin density $\rho_s$ that appears in Eqs. (\ref{dipolar}) and (\ref{fermi}) can be calculated as:
\begin{eqnarray}\label{rhos}
\rho_s(\textbf{r})&=&\frac{1}{Z}\sum_\alpha e^{-\beta E_{\alpha}}\langle \Phi_\alpha^B|c_{\uparrow}^\dag(\textbf{r}) c_{\uparrow}(\textbf{r})-c_{\downarrow}^\dag(\textbf{r}) c_{\downarrow}(\textbf{r}) |\Phi_\alpha^B\rangle \nonumber \\
&=& \sum_i |w(\textbf{r}-\textbf{R}_i)|^2\langle S \rangle
\end{eqnarray}
The second equal sign is obtained by applying Eqs. (\ref{cdag}) and (\ref{S}). Combining Eqs. (\ref{Bint}), (\ref{averageB}) and (\ref{rhos}), we have
\begin{eqnarray}\label{Ahf}
A_{hf}^I&=&\frac{2\mu_0 g_e \mu_B }{3}\frac{\rho_s({\bf{R}}_I)}{\langle S \rangle} \nonumber \\
&=& \frac{2\mu_0 g_e \mu_B }{3}\sum_i |w(\textbf{R}_I-\textbf{R}_i)|^2
\end{eqnarray}
Recall that $I$ denotes the chosen nuclear probe and $i$ denotes the Cu site. Equation (\ref{Ahf}) indicates that $A_{hf}$ only relies on the spatial distribution of the local electrons, while the complicated $f_\alpha^B(\{c_{i\sigma}^\dag\})$ that dictates the many-body state are fully absorbed in $\langle S \rangle$ or equivalently $\chi_S$. When the local electrons are fully polarized, $\sum_i |w(\textbf{R}_I-\textbf{R}_i)|^2$ equals to the total spin density at $\textbf{R}_I$, which rationalizes our calculation proposal.

It is worth reiterating that the justification above relies on (1) the orbital degree of freedom is frozen, and only spin is active to thermal fluctuation and the magnetic field; (2) $B_{ext}$ does not induce spontaneous lattice symmetry breaking, and all the magnetic ions keep equivalent.

The last question is whether $\sum_i |w(\textbf{R}_I-\textbf{R}_i)|^2$ can be evaluated alternatively by explicitly constructing the local orbital $w$, e.g. via Wannierisation, based on a nonmagnetic calculation. In principle, yes, but the issue is that QSL candidates are typically Mott (or charge-transfer) insulators, and within a mean-field theory like DFT, the time-reversal symmetry has to be manually broken in order to open a charge gap. Without the gap opening, the hybridization between the local d-orbital and the anion p-orbitals are significantly overestimated \cite{PRB15DFT_Cu-bdc}.   Furthermore, for the purpose of HFC calculation, the precision of $w$ around the nuclear region is more important than its overall distribution. The former, typically a small quantity, may suffer from large errors during the orbital constructioin process. Therefore, it is most reliable and convenient to calculate the fully-polarized spin density based on the established HFC methodology.

\section{Computational details}

We employ the HFC modules as implemented in WIEN2K~\cite{wien2k} and VASP~\cite{PRB94VASP}.  While both packages are based on DFT, the former is an all-electron (AE) scheme, which is expected to provide the most accurate description around the nucleus; the latter takes advantage of the pseudopotential (PP) approach to reduce the computation cost, but is still able to recover the full wavefunction near the nucleus via the projector-augmented-waves method~\cite{PRB94PAW,1999PRB_PAW_USPP}.

The electron exchange-correlation functional parameterized by Perdew, Burke, and Ernzerhof (PBE)~\cite{PRL96GGA-PBE} is adopted. The strong correlation effect of the Cu-3$d$ orbitals is treated by the DFT+U method~\cite{PRB93SIC,PRB1998LDA+U}, which correctly reproduces the insulating charge gap of Herbertsmithite and Zn-Barlowite. We choose an effective interaction strength $U=6$~eV following the previous works~\cite{PRB15DFT_Cu-bdc,PRB13_valenti_DFT_HS}, which gives a gap size $\sim$ 2 eV. $A_{hf}$ does not sensitively depend on this choice, as long as U is within a reasonable range. The dependence of the results on the $U_\text{eff}$ value is summarized in Tab. I. 

\begin{table}
\caption{Calculated $A_{hf}^{F}$ (kOe/$\mu_\text{B}$), the magnetic moment of the Cu$^{2+}$ ions , and the charge gap $\Delta$ (eV) in Zn-Barlowite Cu$_3$Zn(OH)$_6$FBr, using different methods and $U_\text{eff}$ values.}
\setlength\extrarowheight{3pt}
\begin{ruledtabular}
\begin{tabular}{ccccc}
Method & $U_\text{eff}$ (eV) & $A_{hf}^{F}$ (kOe/$\mu_\text{B}$)  & M$_\text{Cu}$ ($\mu_B$) & $\Delta$ (eV)\\
\hline
\multirow{3}*{AE}
& 4  & 2.522 & 0.732 & 1.40 \\
& 5  & 2.402 & 0.754 & 1.80 \\
& 6  & 2.284 & 0.776 & 2.22\\
\hline
\multirow{3}*{PP}
& 4  & 3.505 & 0.705 & 1.50 \\
& 5  & 3.300 & 0.727 & 1.79 \\
& 6  & 3.120 & 0.750 & 2.08 \\
\end{tabular}
\end{ruledtabular}
\label{table-1}
\end{table}

For the PP-based calculations in VASP, we report the results calculated with an energy cutoff of 450 eV and a $\Gamma$-centered k-point grid. The grid density for unit cell is $8\times8\times6$, while for supercells only the $\Gamma$ point is used. The convergence threshold for self-consistent-field calculations is 10$^{-5}$ eV.

For the full potential augmented plane waves+local orbitals (APW+lo) calculations in WIEN2K, we set the radii of atomic spheres ($R_\text{MT}$) are 1.98, 2.14, 2.18, 2.50, 1.21 and 0.65 a.u. for Cu, Zn F, Br, O, and H respectively. The plane-wave basis of the wave function in the interstitial region are truncated at $K_\text{max}=3.0/min[R_\text{MT}]$.

\section{Benchmark tests}

We apply the proposed calculation method to evaluate $A_{hf}$ of Cl in Herbertsmithite, F in Zn-Barlowite and O in both compounds.  Except for O in Zn-Barlowite, experimental $A_{hf}$ values fitted from the $K$-$\chi_S^{marco}$ slope have been reported. We note that the coefficients in Eqs. (\ref{dipolar}) and (\ref{fermi}) are derived in SI units, and the unit of $A_\textrm{hf}$ is Tesla ($T$).  On the other hand, the commonly-used HFC unit in experiment is $kOe/\mu_B$. The conversion is made by normalizing the total spin density such that the magnetic moment of each  Cu$^{2+}$ is 1 $\mu_B$, and by multiplying a factor of 10 to convert $T$ into $kOe$.

\begin{figure}
\centering
\includegraphics[width=6cm]{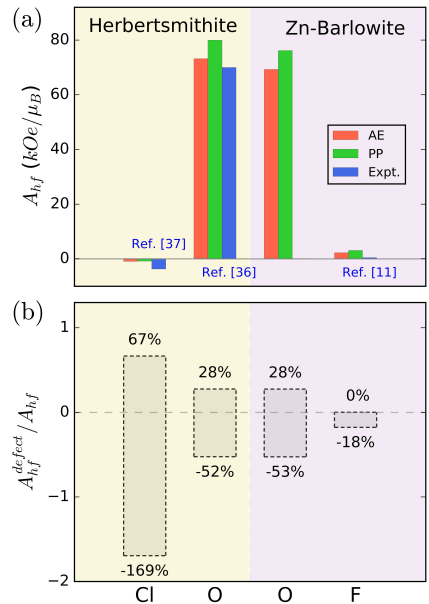}
\caption{(a) Calculated HFC constants of Cl and O in Herbertsmithite, and O and F in Zn-Barlowite, with available NMR experimental data for comparison. ``AE'' and ``PP'' represent all-electron and pseudopotential calculation results respectively. (b) Calculated variation range of HFC when Cu-Zn defects are present. The definition of $A_{hf}^{defect}$ is given in Eq.(\ref{eq:Adefect}).}
\label{fig-1}
\end{figure}

A comparison is made in  Fig.~\ref{fig-1}(a). The overall agreement between the calculation and the experiment is very good. In both compounds, $A_{hf}$ at the O site is at least one order of magnitude larger than that at the halogen site because of the shorter Cu-O distance. Not only the magnitude variation but also the sign of $A_{hf}$ is correctly reproduced by our calculation. In Herbertsmithite, $A_{hf}$ at the Cl site is negative, indicating that the nuclear feels a $B_{int}$ opposite to $B_{ext}$, and thus the larger $\chi_s$ is, the lower the frequency will $K_s$ shift  to. We should point out that when $A_{hf}$ is small, e.g. for the Cl and F cases, the spin density tail at the nuclear site becomes extremely low, so one has to expect a relatively larger error using the experimental value $A_{hf}^{exp}$ as a measure. It appears that the AE method can always reduce the error slightly compared with the PP method, but generally the latter has an adequate accuracy. The inevitable presence of defects in real samples should also have contributed to the differences between the experimental and calculated $A_{hf}$.

 \begin{figure*}
\centering
\includegraphics[width=11.5cm]{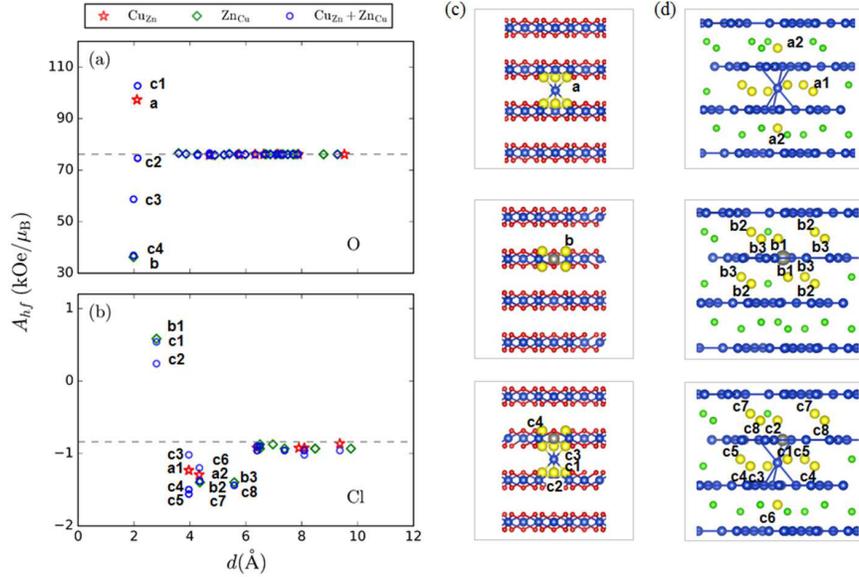}
\caption{Calculated $A_{hf}$ of (a) O and (b) Cl in Herbertsmithite in the presence of three types of Cu-Zn defects. The $A_{hf}$ values vary as a function of distance between the probe nuclei and the defect. For the Cu$_\text{Zn}$+Zn$_\text{Cu}$ pair, the distance is measured depending on which defect atom is closer. The horizontal dash lines indicate the $A_{hf}$ values calculated in the pristine crystal. The strongly perturbed nuclei are marked by letters and numbers, with their positions highlighted in (c) and (d). For clarity, only Cu and the probe nuclei are shown.}
\label{fig-2}
\end{figure*}

\begin{figure*}
\centering
\includegraphics[width=11.5cm]{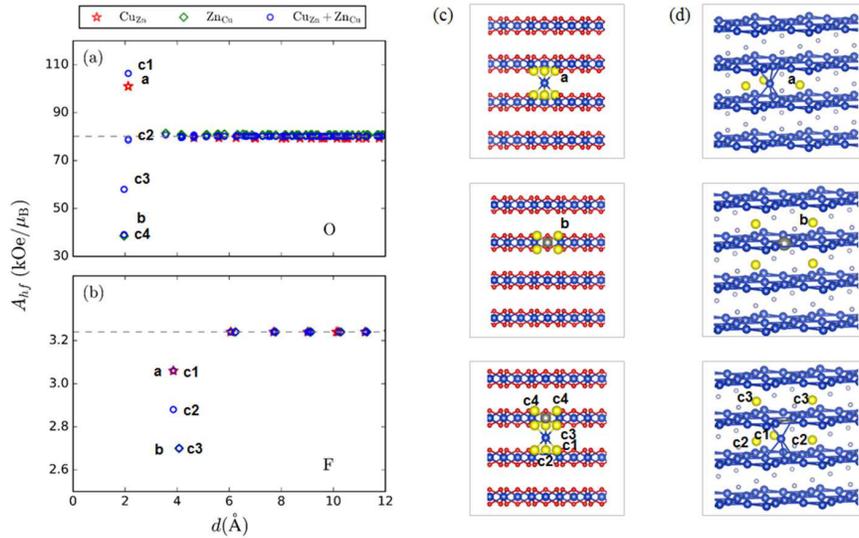}
\caption{Calculated $A_{hf}$ of (a,c) O and (b,d) F in Zn-Barlowite in the presence of three types of Cu-Zn defects. The meaning of the labels is the same as in Fig.~\ref{fig-2} }
 \label{Ahf_Barlo}
\end{figure*}

\section{Analysis of defect effects}

With the benchmark calculation on pristine unit cells, we now proceed to apply this method to defect structures. Considering that the defect simulation is computational demanding, we rely on the less costly pseudopotential method. Three typical types of Cu-Zn defects are created in a large supercell structure:  (a) substituting an inter-kagome Zn with Cu (Cu$_\text{Zn}$), which leads to an extra spin site; (b) substituting a kagome Cu with Zn (Zn$_\text{Cu}$), which results in a spin vacancy; and (c) a Cu$_\text{Zn}$+Zn$_\text{Cu}$ antisite pair.  

\subsection{HFC variation}

Figures~\ref{fig-2}(a) and (b) show $A_{hf}$ in Herbertsmithite as a function of the distance ($d$) between the defect and the selected nuclear sites (O and Cl). Figures~\ref{fig-2}(c) and (d) display the defect structures and highlight the strongly perturbed nuclei. These results reveal remarkable differences between O and Cl: It is clear from Fig.~\ref{fig-2}(a) that $A_{hf}^O(d)$ falls sharply back to the defect-free level beyond 2\text{\AA}. In contrast, $A_{hf}^{Cl}(d)$ has a long tail [Fig.~\ref{fig-2}(b)]. At the O site closest to the defect, an extra spin (Cu$_\text{Zn}$) results in a larger $\rho_s({\bf{R}}_I)$ in Eq. (\ref{fermi}), and thus $A_{hf}^O$ increases; a spin vacancy (Zn$_\text{Cu}$) cuts $A_{hf}^O$  nearly by half; an antisite pair acts in both ways, depending on the O location. These features are intuitive within a classical picture. In contrast, Cu$_\text{Zn}$ reduces $A_{hf}^{Cl}$, suggesting that instead of contributing additive spin density, this inter-kagome spin draws spin density away from Cl. In addition, the Zn$_\text{Cu}$-induced change of $A_{hf}^{Cl}$ exhibits an oscillation with distance, e.g. at the NN site $A_{hf}^{Cl}$ increases, while at the next NN site $A_{hf}^{Cl}$ decreases.  An antisite pair gives rather complicated distribution of $A_{hf}^{Cl}$ [Fig.~\ref{fig-2}(d)].

Figure~\ref{Ahf_Barlo} shows the results of Zn-Barlowite. The behavior of O is very similar to that in Herbertsmithite. Interestingly, F is quite different from Cl. It is clear that F is also a nearsighted probe, and all the simulated defects reduce $A_{hf}$ at the NN F site.

We can define the change of $A_{hf}(d)$ from the defect-free level as a measure of the defect-induced HFC:
\begin{eqnarray} \label{eq:Adefect}
A_{hf}^{defect}(d)=A_{hf}(d)-A_{hf}(d\rightarrow\infty),
\end{eqnarray}
where $A_{hf}(d\rightarrow\infty)$ converges to the $A_{hf}$ values calculated in a pristine unit cell. Figure~\ref{fig-1}(b) summarizes the distribution of  $A_{hf}^{defect}(d)$ with respect to the four nuclear probes. Cl suffers from the largest relative variation, O in between, and F is least affected by the defects.  

Our numerical simulation confirms that O is an excellent probe of the intrinsic kagome physics - the intrinsic $A_{hf}$ is large and the nearsightedness ensures that only a small fraction of O sites is located in an effective magnetic field contaminated by the defects. In contrast, the nonlocal and oscillating $A_{hf}^{defect}$ of Cl in Herbertsmithite inevitably hinders a transparent extraction of $\chi_S^{kagome}(T)$ from the NMR shift.  F in Zn-Barlowite also represents a good local probe. Despite a small coupling to the kagome Cu, F is at the same time less affected by the defects. Its $A_{hf}^{defect}/A_{hf}$ ratio is even smaller than that for O [Fig.~\ref{fig-1}(b)].

We would like to mention that the recent $^{17}$O NMR measurement on single-crystal Herbertsmithite~\cite{Science15NMR_HS} clearly resolved two sets of resonance peaks, one from the O sites in defect-free environment and the other from the O sites closest to the defects. The latter roughly experiences a $A_{hf}$ half of the strength of the former, in agreement with the calculated lower bound of $A^{defect}_{hf}$ as shown in Fig.~\ref{fig-1}(b). In our calculation, this half $A_{hf}$ corresponds to the NN O sites around a Zn$_\text{Cu}$ defect, which has an intuitive explanation as one of the two neighboring Cu ions is missing. This scenario was also presumed in an earlier powder $^{17}$O NMR measurement~\cite{PRL08NMR_HS_intrinsic}, but the new single-crystal experiment~\cite{Science15NMR_HS} showed evidence that this half $A_{hf}$ should be assigned to the NN O sites around a Cu$_\text{Zn}$ defect. The exact type of defects in the samples remains an important question requiring further investigations.

\subsection{Experimental implications} 

\begin{figure}
\centering
\includegraphics[width=7cm]{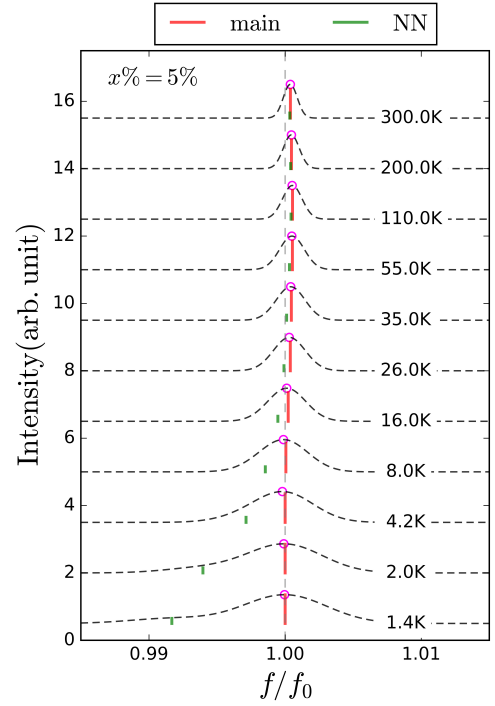}
\caption{ Simulated $^{19}$F NMR spectra in Zn-Barlowite with 5\% Cu$_\text{Zn}$ defects at varying temperature. The red and green bars represent the resonance peak of nuclei far away and nearest to the defect, respectively. The dashed curve serves as a guide for eyes, reflecting that when the defect types become more complex, more satellite peaks exist in between, and merge into a smooth envelop.}
 \label{spectra}
\end{figure}

It is understood that the existence of defects perturbs both $A_{hf}$ and $\chi_S$ in Eq.~(\ref{eq:K}). Cu$_\text{Zn}$ represents the simplest case, which can be considered as a nearly free spin~\cite{PRL08NMR_HS_Imai}, and the NMR shift of a probe nucleus affected by this defect has the form:
\begin{eqnarray}\label{Kdefect}
K(d)&=&A_{hf}\chi_S^{kagome}(T)+A_{hf}^{defect}(d)\chi_S^{defect}(T) \nonumber \\
&=&A_{hf}\chi_S^{kagome}(T)+\frac{A_{hf}^{defect}(d)C}{T},
\end{eqnarray}
where $C$ is the Curie constant. The intrinsic NMR shift $A_{hf}\chi_S^{kagome}(T)$ is thus split into a spectrum of satellite peaks due to the additional defect term. The number of satellite peaks is dictated by the distribution of $A_{hf}^{defect}(d)$, and the peak height is proportional to the population of  the specific nuclear sites. Depending on the sign of $A_{hf}^{defect}(d)$, the satellite peaks can undergo either a blueshift or a redshift. The shift grows rapidly as $T\rightarrow 0$, because of the Curie behavior of the defect susceptibility. When the frequency resolution is low, these satellite peaks simply merge into a broad envelope, and the defect term is responsible for a temperature-dependent envelope width. In powder samples, the half-height-full-width of the broad NMR envelop indeed displays a Curie behavior~\cite{PRL08NMR_HS_Imai}, just like $\chi_S^{macro}$.

It is insightful to consider one concrete example by  plugging in Eq. (\ref{Kdefect}) the calculated $A_{hf}^{defect}$ values. We select F in Zn-Barlowite (Fig. \ref{spectra}), for which a direct comparison can be made against the experimental Fig. 3(a) in Ref~\cite{CPL17NMR_barlowite}  . We adopt the same fitting formula of $\chi_S^{kagome }(T)$ as used in the experimental paper~\cite{CPL17NMR_barlowite}. For each experimental temperature , we calculate the NMR shift at the F sites far from a defect and the F sites nearest to a Cu$_{Zn}$, and mark in Fig. \ref{spectra} with two bars. The bar height is proportional to the population of the corresponding sites. We assume that when the defect type becomes more complex, the other types of F sites in general experience a HFC strength in between these two bars, which merge together into a smooth peak. We can see that the broadening of the NMR peak as temperature drops as observed in the experiment can be naturally explained. Another feature is that the merged envelope has a higher shoulder on the low-frequency side, which arises from the negative sign of $A_{hf}^{defect}$ at the F site.

\section{Perspectives}

In collaboration with the NMR experimentalists, we expect that the input of $A_{hf}^{defect}$ will help formulate a more effective way to separate out the defect signals in the low-T data, pinning down the nature of the long-sought QSL state.

From the computational side, the calculation method proposed here still requires further examination of its general applicability, by applying to a wider range of  QSL candidates. One open question is when the spins in a system have drastically different local susceptibility whether a fully-polarized reference state still works. It also remains to be determined when the spin density around a nuclear site is extremely low,  generally how accurate the numerical results will be.  

Traditionally, quite different from the discovery of other exotic quantum materials, such as topological insulators and semimetals, first-principles method was less involved in the search of QSLs, because rarely their properties could be reliably calculated within the DFT framework. We expect that the HFC predictive potential will stimulate broader interests from the DFT community in this rapidly growing field~\cite{PRB18diamond_Cu}, which in turn may shed some new light into understanding the defect-masked QSL physics.

\section*{Acknowledgements} We would like to acknowledge J.-W. Mei, Z. Li and G.Q. Zheng for stimulating this work. This work is supported by NSFC under Grant No. 11774196 and Tsinghua University Initiative Scientific Research Program. S.Z. is supported by the National Postdoctoral Program for Innovative Talents of China (BX201600091) and the Funding from China Postdoctoral Science Foundation (2017M610858). F.L. acknowledges the support from US-DOE (Grant No. DEFG02-04ER46148). Y.Z. is supported by National Key Research and Development Program of China (No.2016YFA0300202), National Natural Science Foundation of China (No.11774306), 
and the Strategic Priority Research Program of Chinese Academy of Sciences (No. XDB28000000).

\bibliography{ref}
\end{document}